\documentclass[prd,%
preprint,%
tightenlines,showpacs,%
nofootinbib,%
superscriptaddress,%
amsfonts,amsmath,%
floatfix]{revtex4}

\usepackage{hyperref}
\usepackage{graphicx}
\usepackage{xcolor}

\newcommand{\sd}{\mathrm{d}}
\begin{document}

\title{Black holes in self-tuning cubic Horndeski cosmology}

\author{William~T.~Emond}
\email{william.emond@nottingham.ac.uk}

\author{Antoine Leh\'{e}bel}
\email{antoine.lehebel@nottingham.ac.uk}

\author{Paul~M.~Saffin}
\email{paul.saffin@nottingham.ac.uk}

\affiliation{School of Physics and Astronomy, University of Nottingham,\\
University Park, Nottingham NG7 2RD, United Kingdom}

\begin{abstract}
Observations of neutron star mergers in the late Universe have given significant restrictions to the class of viable scalar-tensor theories. In this paper we construct black holes within the ``self-tuning" class of this restricted set, whereby the bare cosmological constant is absorbed by the dynamics of the scalar, giving a lower effective cosmological constant. We use analytic expansions at the singularity, black hole and cosmological horizon, and asymptotic region, coupled with numerical solutions, to find well-behaved black holes that asymptote to the self-tuned de Sitter geometry. The geometry differs from standard general relativity black holes near the horizon, and the scalar field velocity provides a hair for the black holes.
\end{abstract}

\date{\today}

\maketitle

\section{Introduction}

It is a remarkable success of the theory of General Relativity (GR) that it has so far withstood the onslaught of ever more precise astrophysical and cosmological data, and at present remains our best low-energy description of gravity. That is not to say, however, that GR is without its problems, particularly in providing a robust description of the dark sector of the universe. These outstanding issues have motivated research into modifications of GR, the prototypical example being scalar-tensor theory, in which one introduces an additional scalar degree of freedom into the gravitational sector. These are typically described by Horndeski theory \cite{Horndeski:1974wa,Deffayet:2009wt,Deffayet:2009mn,Deffayet:2011gz}, or its more recent generalisation, Degenerate Higher Order Scalar Tensor (DHOST) theory \cite{Langlois:2015cwa,Crisostomi:2016czh,BenAchour:2016fzp}. Such an approach has proven particularly popular in recent years for constructing novel descriptions of dark energy. The recent gravitational wave data \cite{Monitor:2017mdv,Creminelli:2017sry,Sakstein:2017xjx,Ezquiaga:2017ekz,Baker:2017hug,Bettoni:2016mij} has dealt a strong blow to such models, proving that in the late Universe light and gravity propagate at identical speeds, and thus ruling out large regions of the parameter space of these theories. However, a simple subclass of Horndeski Lagrangians passes this test. Some models even provide an interesting self-tuning mechanism. 

Self-tuning is the ability of a model to screen an arbitrary cosmological constant at the level of the action, and possibly to yield an independent, effective cosmological constant at the level of the solutions. Examples of self-tuning towards Minkowski spacetime were first provided in \cite{Charmousis:2011bf} and dubbed `Fab Four' models; the mechanism was later generalized to non-zero effective cosmological constants (e.g. \cite{Appleby:2012rx,Martin-Moruno:2015bda,Charmousis:2014mia}). This is particularly interesting in the context of the large cosmological constant problem; as is the case for any parameter in a quantum field theory, the cosmological constant that we observe should be the sum of a bare contribution and of quantum corrections. However, the bare value of the cosmological constant and the associated quantum corrections must compensate at a level of at least one part in $10^{55}$ \cite{Martin:2012bt} in order to yield the observed value. Self-tuning is thus extremely useful, as a way to trade this huge contribution for an effective cosmological constant that matches the observations. 

On the other hand, with gravitational wave detectors and electromagnetic observations, we have entered an era that makes it possible to explore the strong field regime of gravity. It is thus more important than ever to understand the behaviour of scalar-tensor theories in this sector. Of particular interest is the study of astrophysical black holes, to which current observational and detector experiments are sensitive to. Indeed, the incoming data provides a new testing ground in which to probe the validity of such modifications to GR. Determining black hole solutions within scalar-tensor theories is, however, a complicated endeavour, due to the presence of non-minimal couplings between curvature terms and the additional scalar degree of freedom in the gravity sector. Nevertheless, the problem of embedding black hole solutions in self-tuned universes has already been tackled. Such solutions have been obtained through approximate \cite{Babichev:2012re}, numerical \cite{Babichev:2016fbg} or even exact methods \cite{Babichev:2013cya,Kobayashi:2014eva,Babichev:2016rlq,Babichev:2016kdt}. Star solutions were also obtained numerically in similar contexts \cite{Cisterna:2015yla,Cisterna:2016vdx,Maselli:2016gxk,Ogawa:2019gjc}. Exact black hole solutions were only obtained in the framework of shift-symmetric theories, when the action is invariant under an overall shift of the scalar field across spacetime. In this paper, we also focus on the subclass of shift-symmetric theories; this allows us to maintain tractable calculations, while capturing the higher-derivative effects that are essential to self-tuning.

The structure of the paper is as follows. In sec.~\ref{sec:model}, we introduce our model and the associated field equations. Then, in sec.~\ref{sec:cosmo}, we discuss its self-tuning properties. We present the only self-tuning model left in the class of Horndeski theories that pass the gravitational wave tests, together with an exact cosmological solution. Section \ref{sec:bhs} describes how black holes can be embedded in such a universe, both using analytic expansions and numerical techniques. We conclude in sec.~\ref{sec:discussion}.

\section{A shift-symmetric self-tuning model}
\label{sec:model}

The subset of Horndeski theory that passes the gravitational wave test can be parametrized as \cite{Monitor:2017mdv,Creminelli:2017sry,Sakstein:2017xjx,Ezquiaga:2017ekz,Baker:2017hug,Bettoni:2016mij}
\begin{equation}\label{eq:action}
\int\sd^4 x \sqrt{-g}\Big[K(\phi,X)- G_3(\phi,X)\Box\phi+ G_4(\phi)R - M_{\text{Pl}}^2\Lambda\Big], 
\end{equation}
where $M_{\text{Pl}}$ is the reduced Planck mass, $\Lambda$ is the bare cosmological constant, and $K$, $G_3$ and $G_4$ are a priori arbitrary functions of the scalar field $\phi$ and its kinetic density $X = -\frac{1}{2}\nabla^{\mu}\phi\nabla_{\mu}\phi$. Note that $G_4$ is taken to be a function of $\phi$ alone, in order for the theory to remain consistent with the gravitational wave constraints. 
%The equations of motion for the metric $g_{\mu\nu}$, and scalar field $\phi$ following from this action are:
%
%\begin{subequations}
%\begin{align}\label{eq:eoms metric/scalar}
%	\begin{split}
%	0 = \mathcal{E}^{\mu\nu} &= G_4G^{\mu\nu}- G_{4\phi\phi}\big[\nabla^{\mu}\phi\nabla^{\mu}\phi + 2g^{\mu\nu}X\big] - %G_{4\phi}\big[\nabla^{\mu}\nabla^{\nu}\phi - g^{\mu\nu}\Box\phi\big] 
%\\
%\vphantom{\dfrac{M_{\text{Pl}}^2}{2}}&\quad	+ \nabla^{(\mu}G_3\nabla^{\nu)}\phi + %\frac{1}{2}G_{3X}\Box\phi\nabla^{\mu}\phi\nabla^{\nu}\phi - \frac{1}{2}g^{\mu\nu}\nabla^{\lambda}G_3\nabla_{\lambda}\phi 
%\\
%&\quad-\frac{1}{2}K_X\nabla^{\mu}\phi\nabla^{\nu}\phi - \frac{1}{2}g^{\mu\nu}K + \frac{M_{\text{Pl}}^2}{2}g^{\mu\nu}\Lambda,
%	\end{split}
%	\\
%	0 = \mathcal{E}^\phi &= P^\phi +\nabla_{\mu}J^\mu,
%\end{align}
%\end{subequations}
%  
%where
%
%\begin{align}
%P^\phi &= G_{4\phi}R + \nabla^{\mu}G_{3\phi}\nabla_{\mu}\phi + K_\phi,
%\\
%J^\mu &= - G_{3X}\big[\nabla^{\mu}X + \Box\phi\nabla^{\mu}\phi\big] - 2G_{3\phi}\nabla^{\mu}\phi + K_X\nabla^{\mu}\phi,
%\end{align}
%and the subscripts $\phi$ and $X$ denote partial derivatives with respect to $\phi$ and $X$, respectively (e.g. $G_{3X}=\partial G_3/\partial X$).
The element that is essential to self-tuning is the $X$ dependence in $G_3$. Thus, for tractability of the calculations, we will consider the simpler shift-symmetric subset of \eqref{eq:action}:
	\begin{equation}
	K(\phi,X) = X,\quad G_{3}(\phi,X) = G_3(X),\quad G_4(\phi) = \frac{M_{\text{Pl}}^2}{2}\,,
	\end{equation}
which results in the following action:
\begin{equation}\label{eq:shift symmetric action}
S = \int\sd^4 x \sqrt{-g}\Big[X - G_3(X)\Box\phi  + \frac{M_{\text{Pl}}^2}{2}R - M_{\text{Pl}}^2\Lambda\Big].
\end{equation}
The above action was actually first studied in \cite{Deffayet:2010qz}, where it was dubbed Kinetic Gravity Braiding (KGB). It was later incorporated in the generic framework of covariant Galileons \cite{Deffayet:2009wt,Deffayet:2009mn,Deffayet:2011gz}, which were in turn found to be an equivalent formulation of Horndeski's theory \cite{Horndeski:1974wa}. The associated metric and scalar field equations of motion are then given by:
%(cf.~eq.~\eqref{eq:eoms metric/scalar}):
%
\begin{align}
	\begin{split}
	\mathcal{E}_{(g)}^{\mu\nu} =\dfrac{1}{\sqrt{-g}}\,\dfrac{\delta S}{\delta g_{\mu\nu}} &=G_{3X}\left(\nabla^{(\mu}X\nabla^{\nu)}\phi  +  \frac{1}{2}\Box\phi\nabla^{\mu}\phi\nabla^{\nu}\phi - \frac{1}{2}g^{\mu\nu}\nabla_{\lambda}X\nabla^{\lambda}\phi\right)
	\\ 
	&\quad +\frac{M_{\text{Pl}}^2}{2}(G^{\mu\nu} + g^{\mu\nu}\Lambda)  - \frac{1}{2}\nabla^{\mu}\phi\nabla^{\nu}\phi - \frac{1}{2}g^{\mu\nu}X = 0,
		\label{eq:metric eom}
	\end{split}
	\end{align}
	\begin{equation}
	\mathcal{E}_{(\phi)} =\dfrac{1}{\sqrt{-g}}\,\dfrac{\delta S}{\delta \phi} = \nabla_{\mu}J^\mu = 0 ,
	\label{eq:conservation eq}
	\end{equation}
where 
\begin{equation}
\label{eq:conserved current}
J^\mu = \nabla^\mu\phi - G_{3X}(\nabla^\mu X + \Box\phi\nabla^\mu\phi)
\end{equation}
is the conserved Noether current associated to the shift-symmetry, and the subscript $X$ denotes a partial derivative with respect to $X$ (e.g. $G_{3X}=\partial G_3/\partial X$).

\section{Cosmological solutions to the field equations}
\label{sec:cosmo}

Let us first study the cosmological solutions of the model, within which we aim to embed black holes. Any black hole solution should match this behaviour asymptotically. We start with the Friedmann Lema\^{i}tre Robertson Walker (FLRW) geometry:
\begin{equation}\label{eq:flrw}
	\sd s^2 =  - \sd \tau^2 + a^2(\tau)\sd\mathbf{x}^2,
\end{equation}
where $\tau$ is the cosmic time, $a(\tau)$ is the cosmological scale factor, and we adopt the metric signature $(-+++)$ throughout. The scalar field is also assumed to be spatially homogeneous at this scale, i.e. $\phi=\phi(\tau)$. Given this, the metric and scalar equations of motion, eqs.~\eqref{eq:metric eom}-\eqref{eq:conservation eq}, are 
\begin{subequations}\label{eq:cosmological eoms}
	\begin{flalign}
		0 &= 6M_{\text{Pl}}^2H^2 - 6H\dot{\phi}^3G_{3X} - \dot{\phi}^2 - 2M_{\text{Pl}}^2\Lambda,\label{eq:Friedmann constraint}
		\\
		0 &= 2M_{\text{Pl}}^2\Lambda -4M_{\text{Pl}}^2\dot{H} - 6M_{\text{Pl}}^2H^2 +2\dot{\phi}^2\ddot{\phi}G_{3X} - \dot{\phi}^2, \label{eq:Hubble eom}
		\\
		0 &= 3H\dot\phi(1+3G_{3X}H\dot\phi)+3G_{3X}\dot\phi^2\dot{H}+ \big[1 + 3H\dot{\phi}(2G_{3X} + G_{3XX}\dot\phi^2)\big]\ddot{\phi}, \label{eq:cosmo scalar eom}
 	\end{flalign}
\end{subequations}
where a dot denotes a derivative with respect to cosmic time $\tau$, $H=\dot{a}/a$ is the Hubble parameter, and $G_3$ depends implicitly on $\dot\phi$ through $X=\dot{\phi}^2/2$. We now fix the geometry to be de Sitter, i.e. $\dot{H}=0$, $H = H_0$. It immediately follows from eq.~\eqref{eq:Friedmann constraint} that $\dot{\phi}$ is also constant. The $G_3$ dependence then drops from eq.~\eqref{eq:Hubble eom}, from which we can determine that $\dot{\phi}$ is given by
\begin{equation}\label{eq:phidot sol}
	\dot{\phi}_0 = \pm\sqrt{2}M_{\text{Pl}}\sqrt{\Lambda- 3H_0^2}.
\end{equation}
Once substituted in eq.~\eqref{eq:Friedmann constraint}, one realises that $H_0$ will always depend on $\Lambda$, unless $\dot\phi G_{3X}$ is itself insensitive to the value of $\dot\phi$. Thus, to ensure self-tuning, i.e. that the cosmological properties of the metric do not depend on the bare cosmological constant, we set
\begin{equation}\label{eq:G3}
G_3(X) = - \frac{1}{3}\,\frac{\sqrt{2X}}{\mu} = -\frac{1}{3}\,\frac{\dot{\phi}}{\mu}\,,
\end{equation}
where $\mathcal{\mu}$ is a priori an arbitrary parameter in the action~\eqref{eq:action}, with mass dimension $[\mu]=1$. Interestingly, the particular choice of $G_3$ in eq.~\eqref{eq:G3} corresponds to a constant diffusion coefficient in the imperfect fluid interpretation of the KGB model \cite{Pujolas:2011he}. The on-shell field equations then impose that $H_0=\mu$, such that the value of the Hubble parameter is independent of the cosmological constant, i.e. the model admits self-tuning solutions. Having to introduce the current Hubble scale of $\mu\sim10^{-33}$~eV directly at the level of the action can appear unnatural. However, it is counterbalanced by the advantage that $H_0$ is completely independent on $\Lambda$; thus, the tuning will survive phase transitions, that are otherwise extremely problematic in the context of the standard model \cite{Martin:2012bt}.

We have established the cosmology to which we will match black hole solutions. We now need to consider how we can achieve such a matching. We will look for static and spherically symmetric black hole solutions. Thus, we will use the following system of coordinates:
\begin{equation}\label{eq:static metric}
	\sd s^2 = - f(r)\,\sd t^2 + \frac{1}{g(r)}\,\sd r^2 + r^2\sd\Omega^2 ,
\end{equation}
where $\sd\Omega^2 = \sd\theta^2 + \sin^2\theta\,\sd\varphi^2$, and $f(r)$ and $g(r)$ are a priori arbitrary functions to be determined by the field equations\footnote{Note that unlike in GR, the functions $f(r)$ and $g(r)$ do not have to be equal on astrophysical scales, and indeed, they often are not for scalar-tensor theories of gravity.}. The de Sitter metric, eq.~\eqref{eq:flrw} with $a(\tau)=e^{H_0\tau}$, can actually be matched to its well-known static patch through the following invertible change of coordinates \cite{Babichev:2012re}:
\begin{subequations}\label{eq:FLRW coord transform}
	\begin{align}
		\tau &= t + \dfrac{1}{2H_0} \ln\left[1  - (H_0 r)^2\right],
		\label{eq:taut}
		\\
		\rho &= \dfrac{e^{-Ht}}{\sqrt{1-(H_0r)^2}} \,r.
	\end{align}
\end{subequations}
%\begin{subequations}\label{eq:FLRW coord transform}
%	\begin{align}
%		t &= \tau - \dfrac{1}{2H_0} \ln\Big[1  - \Big(H_0 e^{H_0\tau}\rho\Big)^2\Big],
%		\\
%		r &= e^{H_0\tau}\rho.
%	\end{align}
%\end{subequations}
%
Under this change of coordinate, the metric \eqref{eq:flrw} with $a(\tau)=e^{H_0\tau}$ is mapped to 
\begin{equation}
	f(r) = g(r) = 1 - (H_0r)^2.
\end{equation}

\section{Static black hole solutions}
\label{sec:bhs}

\subsection{Ansatz and field equations}

Having determined how a black hole solution within this shift-symmetric model should behave asymptotically, i.e. on cosmological scales, we shall now move on to determine the details of such solutions. For simplicity, we shall consider static and spherically symmetric metric configurations, as given by the ansatz \eqref{eq:static metric}. The scalar field, on the other hand, is taken to depend not only on the radial coordinate $r$, but also linearly on time. This emerges naturally from the combination of eqs.~\eqref{eq:phidot sol}-\eqref{eq:taut}, and was shown to be fully consistent in \cite{Babichev:2015rva}. The stress-energy tensor associated with $\phi$ only involves derivatives of the scalar field; thus it is itself static and one can consistently require a static geometry\footnote{This is certainly restrictive, and dynamical branches of solutions should exist. However, no such solution has been obtained in the literature so far, due to technical difficulty mostly.}. Furthermore, although a naive counting of the field equations and free functions tells us that the system seems overconstrained, ref.~\cite{Babichev:2015rva} proved this is not the case; the $(tr)$ component of the metric equation is always proportional to the radial component of the current, $J^r$. This allows us to look for solutions with the following ansatz for $\phi$:
\begin{equation}\label{eq:spherical symmetry ansatz}
	\phi(t,r) = qt + \int^r\sd r'\,\frac{\chi(r')}{f(r')}\,,
\end{equation}
where $q=\dot\phi$ is a free, non-zero constant for now, and $\chi(r)$ is an arbitrary function to be determined by the field equations. The precise form of the second term on the right hand side is chosen for convenience. Notably, the $f(r)$ denominator allows us to absorb the diverging contribution of the $r$-dependent part of $\phi$ when approaching the black hole horizon. 
%The corresponding kinetic density $X$ is therefore independent of $t$, given by 
%
%\begin{equation}
%X(r) = \frac{q^2}{2f(r)}\Big[1 - \frac{g(r)}{f(r)}\Big(\frac{\chi(r)}{q}\Big)^2\Big].
%\end{equation}
% 
Let us now write down the independent field equations. We will work with the $(tr)$ component of the metric field equations. Additionally, we use specific linear combinations of the $(tt)$ and $(tr)$ field equations on one hand, and $(rr)$ and $(tr)$ field equations on the other. Due to spherical symmetry, $\smash{\mathcal{E}_{(g)}^{\theta\theta}=\sin^2\theta\,\mathcal{E}_{(g)}^{\varphi\varphi}}$, and $\smash{\mathcal{E}_{(g)}^{\theta\theta}}$ itself can be deduced from the previous three equations, due to the diffeomorphism invariance identity \smash{$\mathcal{E}_{(\phi)}\nabla^\nu\phi+2\nabla_{\mu}\mathcal{E}_{(g)}^{\mu\nu}=0$}. The $(tr)$ metric equation reads (up to an overall non-zero factor):
\begin{equation}
\label{eq: psi constraint from scalar eom}
\frac{\chi}{q} + \frac{q}{2f}G_{3X}\Big[f' - \frac{g}{r^4f}\big(r^4f)'\Big(\frac{\chi}{q}\Big)^2\Big] = 0,
\end{equation}
with
\begin{equation}
G_{3X} = -\frac{f}{3q\mu}\Big[f - g\Big(\frac{\chi}{q}\Big)^2\Big]^{-1/2},
\end{equation}
Then, the $(rr)$-$(tr)$ and $(tt)$-$(tr)$ combinations read respectively:
	\begin{align}
	0 &= M_{\text{Pl}}^2\,\frac{g\big(rf\big)' + r^2f\Lambda -f}{2r^2f} - \frac{q^2}{4f}\Big[1 - \frac{g}{f}\Big(\frac{\chi}{q}\Big)^2\Big], \label{eq:Err}
	\\
	\begin{split}
	0 &= M_{\text{Pl}}^2\,\frac{\big(rg\big)' + r^2\Lambda - 1}{2r^2} - \frac{q^2}{2r^2f}\sqrt{\frac{g}{f}}G_{3X}\Big[1 - \frac{g}{f}\Big(\frac{\chi}{q}\Big)^2\Big]\Big(r^2\sqrt{\frac{g}{f}}\chi\Big)' 
	\\
	&\quad+ \frac{q^2}{4f}\Big[1 - \frac{g}{f}\Big(\frac{\chi}{q}\Big)^2\Big]. 
	\label{eq:Ett}
	\end{split}
	\end{align}
Note that the $G_3$ dependence drops out from the $(rr)$-$(tr)$ combination, eq.~\eqref{eq:Err}. It is not possible to solve this system fully analytically, and therefore, we will ultimately have to appeal to numerical techniques. The difficulty of finding exact black hole solutions for cubic Horndeski models was already noted in \cite{Babichev:2016fbg}. In general, it appears that no exact black hole solution is known for Horndeski models that do not possess the reflection symmetry $\phi\to-\phi$, particularly the simplest cubic and quintic models. Indeed, typically the scalar field equation can be integrated once, but it then becomes a high-order algebraic equation for $\phi'$. Exact solutions are known only for models that possess both shift and reflection symmetry \cite{Babichev:2013cya,Kobayashi:2014eva,Babichev:2017guv} (apart from standard scalar-tensor theories). In order to integrate the system numerically, it proves convenient to introduce a set of dimensionless quantities. To this end, we note that the Lagrangian contains three dimensionful parameters: $M_{\text{Pl}}$, $\mu$ and $\Lambda$, along with the scalar field velocity $q$, introduced in the ansatz for $\phi$. Let us further introduce a length scale $r_0$, which physically corresponds to the event horizon radius of the black hole, such that we can define a dimensionless radius $x=r/r_0$. Then, given the parameters of the Lagrangian, the scalar field velocity $q$ and the length scale $r_0$, we can express the field equations using only the following three dimensionless constants:
\begin{equation}
\beta_1 = \frac{1}{r^2_0\mu^2}\,, \quad \beta_2 = \frac{r_0^2q^2}{M_{\text{Pl}}^2} \,, \quad \beta_3 = r_0^2\Lambda.
\label{eq:betas}
\end{equation}
Equations \eqref{eq: psi constraint from scalar eom}, \eqref{eq:Err} and \eqref{eq:Ett} can then be rewritten:
	\begin{align}
	0 &= 6x^4\frac{\chi}{q} - \sqrt{\beta_1}\Big[x^4\frac{\sd f}{\sd x} - \frac{g}{f}\frac{\sd}{\sd x}\big(x^4f)\Big(\frac{\chi}{q}\Big)^2\Big]\Big[f - g\Big(\frac{\chi}{q}\Big)^2\Big]^{-1/2}, 
	\label{eq:alpha 1 scalar eom} 
	\\
	0 &= 2\Big[g\frac{\sd}{\sd x}\big(xf\big) + \beta_3x^2f- f\Big] - \beta_2x^2\Big[1 - \frac{g}{f}\Big(\frac{\chi}{q}\Big)^2\Big] ,
	\label{eq:alpha 1 Err}
	\\
	\begin{split}
	0 &= 6\Big[\frac{\sd}{\sd x}\big(xg\big) + \beta_3x^2 -  1 \Big] + \frac{3\beta_2x^2}{f}\Big[1 + \frac{g}{f}\Big(\frac{\chi}{q}\Big)^2\Big] 
	\\
	&\quad + \frac{2\sqrt{\beta_1g}\beta_2}{f}\Big[1 -\frac{g}{f}\Big(\frac{\chi}{q}\Big)^2\Big]^{1/2}\frac{\sd}{\sd x}\Big(x^2\sqrt{\frac{g}{f}}\frac{\chi}{q}\Big).
	\label{eq:alpha 1 Ett}
	\end{split}
	\end{align}
These will be the equations that we integrate numerically.

\subsection{Analytical approximations in the small \& large $r$ limits}
\label{sec:rexpansion}

Before solving eqs.~\eqref{eq:alpha 1 scalar eom}, \eqref{eq:alpha 1 Err} and \eqref{eq:alpha 1 Ett} numerically, we shall study the asymptotic behaviour of $f$, $g$ and $\chi$ in the small and large $r$ limits. Solving the system of equations near the origin $r= 0$, we find that 
\begin{subequations}
\label{eq:smallrexp}
	\begin{flalign}
		f(r) &\underset{r\to0}{=} - \frac{a}{r^4} + \mathcal{O}(r^{-3}),
		\\
		g(r) &\underset{r\to0}{=} - \frac{1}{3} -  br  + \mathcal{O}(r^2),
		\\
		\chi(r) &\underset{r\to0}{=} - \frac{c}{r^{9/2}} + \mathcal{O}(r^{-7/2}),
	\end{flalign}
\end{subequations}
where $a$, $b$ and $c$ are fixed by the field equations \eqref{eq: psi constraint from scalar eom}, \eqref{eq:Err} and \eqref{eq:Ett} (we omit their exact expressions here in the interest of succinctness). We note that, as was found in~\cite{Babichev:2016fbg}, the $g(r)$ component of the metric is finite at the origin, unlike in GR. It is possible that this is a generic feature of any cubic Horndeski model for which such black hole solutions exist. 

Let us now consider the large $r$ asymptotic behaviour of the solutions. We assume that, as $r\to \infty$, the solution has an analytic expansion in powers of $1/r$.
%
%\begin{equation}
%		f(r) = \sum_{n=-2}^{\infty}\frac{a_n}{r^n} \;,\qquad g(r) \ = \ \sum_{n=-2}^{\infty}\frac{b_n}{r^n} \;,\qquad \chi(r) \ = \ \sum_{n=-1}^{\infty}\frac{c_n}{r^n} \;,
%\end{equation}
%
%Inserting these expansions into eqs.~\eqref{eq: psi constraint from scalar eom} and~\eqref{eq:metric eoms}, and solving at each order in $1/r$, 
We find that the asymptotic behaviour of the solution is given by:
\begin{subequations}\label{eq:large r sol}
	\begin{flalign}
		f(r) &\underset{r\to\infty}{=} \Big(\frac{q}{q_0}\Big)^2\left[1 - (H_0r)^2\right] + \mathcal{O}(r^{-1}),
		\\
		g(r) &\underset{r\to\infty}{=} 1 - (H_0r)^2 + \mathcal{O}(r^{-1}),
		\\
		\chi(r) &\underset{r\to\infty}{=} - \dfrac{\mu q^2}{q_\mathrm{0}} r  + \mathcal{O}(r^{-1}),
	\end{flalign}
\end{subequations}
where $q_0=\dot\phi_0$ is given by eq.~\eqref{eq:phidot sol}. Note that, at this point, the velocity parameter $q$ remains arbitrary, and may not necessarily coincide with $q_0$, i.e., the velocity parameter corresponding to the cosmological solution. However, it is interesting to check whether the scalar field in the asymptotic solution~\eqref{eq:large r sol} is homogeneous in the set of cosmological coordinates $(\tau,\rho)$. From eqs.~\eqref{eq:spherical symmetry ansatz} and~\eqref{eq:large r sol}, it follows that 
\begin{equation}
	\phi(t,r)\underset{r\to\infty}{=} qt +\frac{\mu q}{H_0^2}\,\text{ln}(r) +\mathcal{O}(r^{-2}).
\end{equation}
Similarly, using the coordinate transformation~\eqref{eq:FLRW coord transform}, $\phi$ can be expanded as
\begin{equation}
	\phi(\tau,\rho)\underset{\rho\to\infty}{=}q\tau +\dfrac{q}{\mu}\Big(\dfrac{q_0}{q}-1\Big)\text{ln}(\rho) +\mathcal{O}(\rho^{-2})
\end{equation}
in cosmological coordinates. Thus, we see that $\phi$ is asymptotically homogeneous only when the local velocity $q$ matches the asymptotic cosmological value, as was already the case in \cite{Babichev:2016fbg}. Whether the solutions with $q\neq q_0$ are physically acceptable is arguable, since only the gradient of $\phi$ is physically relevant, and the latter always decays as $1/\rho$. In our numerical analysis, still, we chose to impose that $q=q_0$ to ensure a safe cosmological behaviour.

Finally, let us comment briefly on the fate of the scalar field at the (black hole and cosmological) horizons. It is well known \cite{Babichev:2013cya} that in the system of coordinates \eqref{eq:static metric}, the $r$-dependent part of $\phi$ diverges at both horizons. However, this is a coordinate effect, and this divergence can be taken care of by working in adapted Eddington-Finkelstein coordinates. Note that it is not in general possible to absorb the divergence for both horizons. However, recent works have shown that generalising to the rotating case can cure the divergence of $\phi$ on both horizons for this type of solution \cite{Charmousis:2019vnf}.

\subsection{Numerical integration of the field equations}

\begin{figure}[ht]
\centering
\includegraphics[width=.8\textwidth]{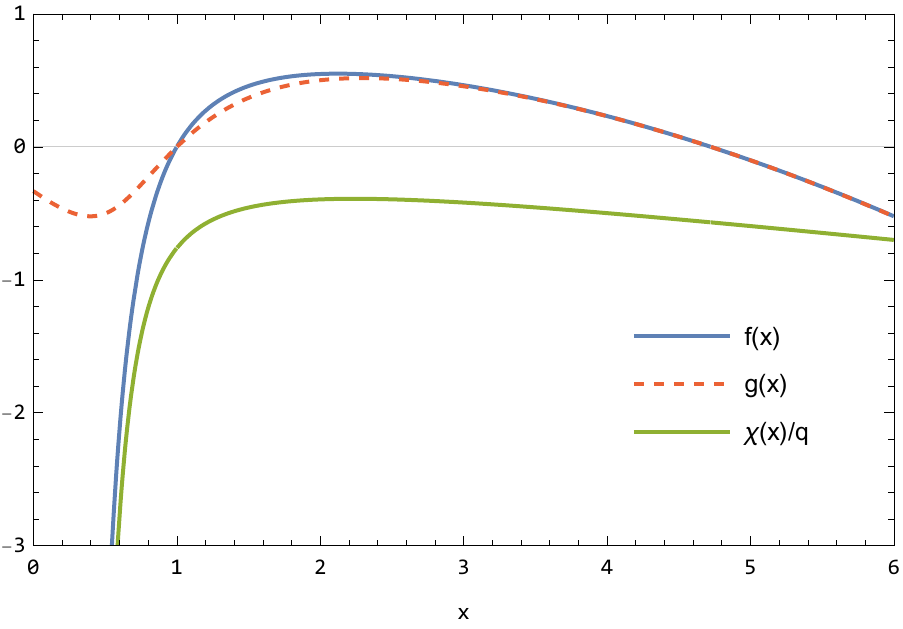}
\caption{Typical solution of the field equations. In this plot, the parameters are chosen to be $\beta_1=100$, $\beta_2=1.76$, $\beta_3=1$. This corresponds to a modest hierarchy of $\beta_1\beta_3=10^2$ between the bare and effective cosmological constants. The ratio can be increased, but we kept these scales for clarity of the plots.}
\label{fig:typicsol}
\end{figure}

\begin{figure}[ht]
\centering
\includegraphics[width=\textwidth]{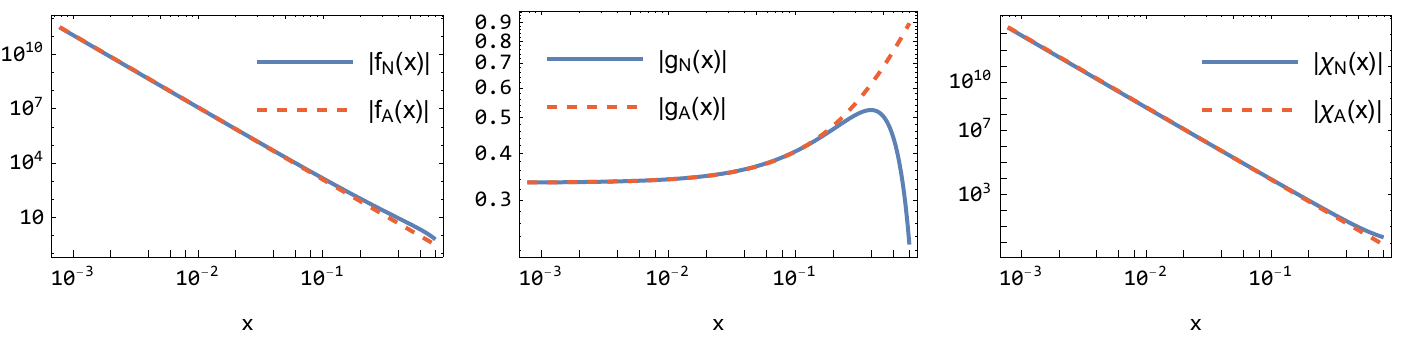}
\caption{Logarithmic plot of the metric and scalar field functions at small radius, together with the analytic expansion found in eq.~\eqref{eq:smallrexp}. A subscript N in the legend indicates that the plotted function is obtained through numerical simulation, while a subscript A indicates that we plot the corresponding analytic expansion. The parameters are the same as in Fig.~\ref{fig:typicsol}.}
\label{fig:smallr}
\end{figure}

\begin{figure}[ht]
\centering
\includegraphics[width=\textwidth]{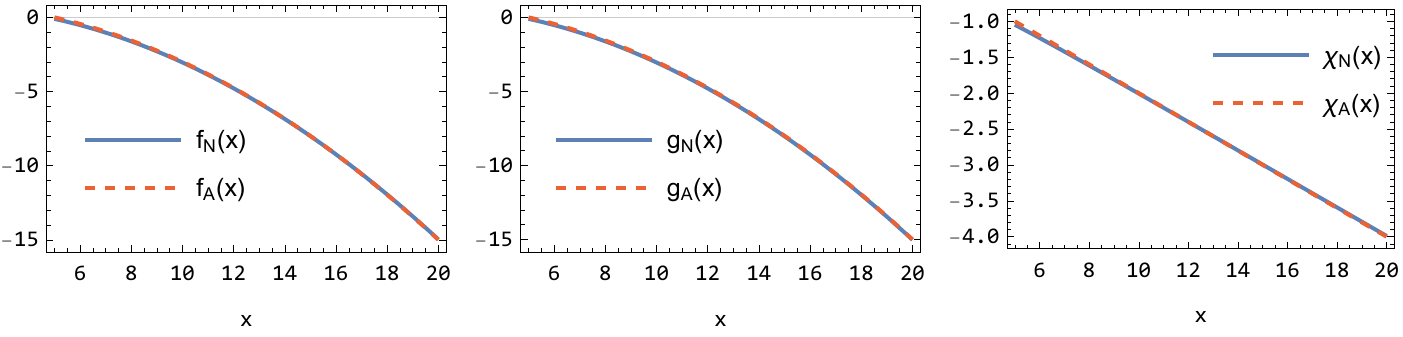}
\caption{The metric and scalar field functions at large radius, together with the asymptotic expansion of eq.~\eqref{eq:large r sol}. Again, the parameters are the same as in Fig.~\ref{fig:typicsol}, and the convention in the legend is the same as in Fig.~\ref{fig:smallr}.}
\label{fig:larger}
\end{figure}

Having studied the asymptotics of the solutions, we shall now proceed to carry out a numerical integration of this system. Note that eqs.~\eqref{eq:alpha 1 scalar eom} and~\eqref{eq:alpha 1 Err} are algebraic equations in $g$ and $\chi$, and can thus, in principle, be solved to express them in terms of $f$ and $\sd f/\sd x$. Upon substitution into eq.~\eqref{eq:alpha 1 Ett}, we are then left with a second-order ordinary differential equation (ODE) for $f$. In practice, it turns out the system is more readily solved numerically by first solving eq.~\eqref{eq:alpha 1 Err} to determine $\chi$ in terms of $f$, $\sd f/\sd x$ and $g$, and then solving the remaining two equations as a system of ODEs for $f$ and $g$. This requires us to specify several boundary conditions in order to obtain the unique solution. By imposing that both $f$ and $g$ vanish at $x=1$, we define the black hole horizon to be located at this point (corresponding to $r=r_0$ as mentioned earlier). Moreover, having imposed $f\vert_{x=1}=0$, we can then expand eq.~\eqref{eq:alpha 1 scalar eom} around $x=1$ to determine \smash{$\frac{\sd f}{\sd x}\vert_{x=1}$} in terms of \smash{$\frac{\sd g}{\sd x}\vert_{x=1}$} (or vice versa). This leaves us with one remaining boundary condition that we are free to choose arbitrarily, although ultimately the choice is dictated by requiring that at large $x$, the solution has the desired asymptotic cosmological behaviour. Thus, for given parameters, we use the shooting method to pick the value of \smash{$\frac{\sd g}{\sd x}\vert_{x=1}$} such that \smash{$f\underset{x\to\infty}{\sim}g$}. We start the numerical integration slightly outside the black hole horizon. Then, the code breaks down when approaching both the black hole and cosmological horizons. It is however very easy to perform a series expansion around these points, and to restart the numerical integration on the other side of the horizon(s). Thus, we are able to compute the solution over the full range $0<x<+\infty$. A typical solution is shown in Fig.~\ref{fig:typicsol}.
More detailed plots of the small and large $r$ behaviour are shown in Figs.~\ref{fig:smallr} and \ref{fig:larger} respectively. The numerical solutions agrees with the analytic expansion presented in sec.~\ref{sec:rexpansion}. As mentioned in sec.~\ref{sec:rexpansion}, we stick to $q=q_0$ in order to ensure the good asymptotic behavior. In terms of dimensionless parameters, this translates as
\begin{equation}
q=q_0 \Leftrightarrow \beta_2=\dfrac{2}{\beta_1}(\beta_1\beta_3-3).
\end{equation}

\section{Discussion}
\label{sec:discussion}

We have considered the class of Lagrangians within Horndeski theory for which light and gravity travel with identical speeds, as motivated by observational data in the late Universe \cite{Monitor:2017mdv,Creminelli:2017sry,Sakstein:2017xjx,Ezquiaga:2017ekz,Baker:2017hug,Bettoni:2016mij}. We restricted ourselves to shift-symmetric models, in order to keep the essential higher derivative effects while handling reasonably tractable equations. In this context, only one model exhibits a fully self-tuning behaviour, allowing an exact de Sitter solution fully independent of the bare cosmological constant. We discussed how black hole solutions can be matched with these cosmological asymptotics using an analytic expansion. We then explicitly built such solutions using a numerical shooting method. The numerical solutions agree with the analytic expansions, both at large radius and close to the central singularity. We notice that, at the centre of the black hole, one of the metric components remains finite, a behaviour that was already found in \cite{Babichev:2016fbg} for a similar model. 

It is interesting to note that the Lagrangian \eqref{eq:shift symmetric action}-\eqref{eq:G3} falls in the well-tempered category \cite{Appleby:2018yci,Emond:2018fvv}. The well-tempered proposal is somehow similar to self-tuning, but it aims at treating dark energy separately. The motivation of the well-tempered model is to resolve a problem of the previous self-tuning ``fab-four" model \cite{Charmousis:2011bf}, whereby the model's ability to absorb a huge cosmological constant came at the price of it greatly affecting any matter/radiation content as well.
%
%the following: a self-tuning field that is able to eat up a huge cosmological constant might absorb any other matter content as well. 
%
A consistent cosmology requires, in particular, a radiation and a matter dominated era, so one should be careful to allow such an era in the cosmological history. In the well-tempered set-up, dark energy is on a special footing due to its particular equation of state $P+\rho=0$. This allows for the scalar field to only self-tune for constant energy densities, such that radiation and matter dominated eras seem to be generically present. However, in this paper, we focused on exact de Sitter solutions for the asymptotics. In the well-tempered context, exact de Sitter solutions exist only for a certain relation of the type $H\simeq\mu\simeq\sqrt{\Lambda}$, which we obviously want to avoid. Thus, although the model we study is in the well-tempered class, we do not focus on a well-tempered branch but on a self-tuning one. Still, it seems that well-tempered models possess solutions that approach $H\simeq\mu\ll\sqrt{\Lambda}$ asymptotically \cite{Emond:2018fvv,Appleby:2018yci}. An interesting extension of our work would be to study the well-tempered branches. This however comes with the extra difficulty that the solution is only approximately de Sitter at the cosmological level. It cannot be matched to a static patch, and one would likely have to consider non-static metrics.

We also note, following ref.~\cite{Babichev:2016kdt}, that local gravity tests (such as Solar System tests) are likely to fail for our model if one wants to tune too large a value of the bare cosmological constant. This is true even when non-linear effects, usually responsible for a Vainshtein screening, are taken into account. Perhaps this issue can be addressed by introducing non-shift symmetric terms in the $K$ or $G_3$ functions, relying on lower derivative screening mechanisms. 

Finally, the question of the stability could be examined in order to carry further the works presented here. Notably, the scalar field perturbations will propagate in an effective metric that differs from the spacetime metric. This metric can be found in \cite{Deffayet:2010qz}. One should check whether this metric is always Lorentzian, and if the associated causal cone is compatible with the causal cone of the spacetime metric, in the sense of refs.~\cite{Babichev:2017lmw,Babichev:2018uiw}. Another possible extension of our work would be to consider the thermodynamics of the black hole solutions presented here. The temperature and entropy could be computed along the lines of, e.g., \cite{Mignemi:1992nt}. There are no conceptual difficulties in applying such a procedure; however, since no analytic solution is available in our case, this analysis would require a full numerical sampling over the parameter space of the model, eq. \eqref{eq:betas}, which is beyond the scope of this paper.

\acknowledgments{We thank Christos Charmousis for interesting discussions. A.L. and W.T.E. are funded by an STFC Consolidated Grant.}

\bibliographystyle{unsrt}
\bibliography{ST_BH_bibliography}{}

\end{document}